\documentclass{vgtc}         

\ifpdf
  \pdfoutput=1\relax                   
  \usepackage{graphicx}                
  \DeclareGraphicsExtensions{.pdf,.png,.jpg,.jpeg} 
\else
  \usepackage{graphicx}                
  \DeclareGraphicsExtensions{.eps}     
\fi%
\usepackage{multirow}
\graphicspath{{../figures/}{pictures/}{images/}{./}} 
\usepackage{float}
\usepackage{microtype}                 
\PassOptionsToPackage{warn}{textcomp}  
\usepackage{textcomp}                  
\usepackage{mathptmx}                  
\usepackage{times}                     
\usepackage{cite}                      
\usepackage{tabu}                      
\usepackage{booktabs}                  
\usepackage{hyperref}
\usepackage{amssymb}
\usepackage{mathtools}

\onlineid{1001}

\vgtccategory{Research}
\vgtcpapertype{Application}

\title{Topological Analysis of
Ensembles of Hydrodynamic Turbulent Flows\\
An Experimental Study}

\author{Florent Nauleau \\
\scriptsize CEA \and Fabien Vivodtzev \\
\scriptsize CEA
\and Thibault Bridel-Bertomeu\thanks{\{firstname.lastname\}@cea.fr} \\
\scriptsize CEA
\and H\'elo\"ise Beaugendre\thanks{\{firstname.lastname\}@math.u-bordeaux.fr}\\
\scriptsize University of Bordeaux
\and Julien Tierny\thanks{\{firstname.lastname\}@sorbonne-universite.fr}\\
\scriptsize CNRS}
\authorfooter{
\item
 Florent Nauleau, Thibault Bridel-Bertomeu and Fabien Vivodtzev are with the
CEA. E-mail: \{firstname.lastname\}@cea.fr.
\item
 H\'elo\"ise Beaugendre is with University of Bordeaux and Bordeaux INP. E-mail:
firstname.lastname@math.u-bordeaux.fr.
\item
Julien Tierny is with Sorbonne Universit\'e and CNRS. E-mail:
firstname.lastname@sorbonne-universite.fr.
}

\shortauthortitle{Biv \MakeLowercase{\textit{et al.}}: Global Illumination for Fun and Profit}

\abstract{
This application paper presents a comprehensive experimental evaluation of the
suitability of Topological Data Analysis (TDA)
for the quantitative comparison of turbulent flows. Specifically, our
study documents the usage of the persistence diagram of the maxima of flow
enstrophy (an established vorticity indicator), for the topological representation of 180 ensemble members, generated
by a coarse sampling of the parameter space of five numerical solvers.
We document five main hypotheses reported by domain experts, describing their
expectations regarding the variability of the flows generated by the distinct
solver configurations. We contribute three evaluation protocols to assess the
validation of the above hypotheses by two comparison measures:
\emph{(i)} a standard distance used in scientific imaging (the $L_2$ norm) and
\emph{(ii)} an established topological distance between persistence diagrams
(the $L_2$-Wasserstein metric).
Extensive experiments on the input ensemble demonstrate
the
superiority of the topological distance (\emph{ii}) to report as close to each
other flows which are expected to be similar by domain experts, due to the
configuration
of their vortices.
Overall,
the insights reported by
our study bring an experimental evidence of the suitability
of TDA for representing and comparing turbulent flows, thereby providing
to the fluid dynamics community
confidence for its usage in future work.
Also, our flow data and evaluation protocols provide to the TDA community an
application-approved benchmark for the evaluation and design of further
topological distances.}
\teaser{
  \centering
  \includegraphics[width=\figureShrink\linewidth]{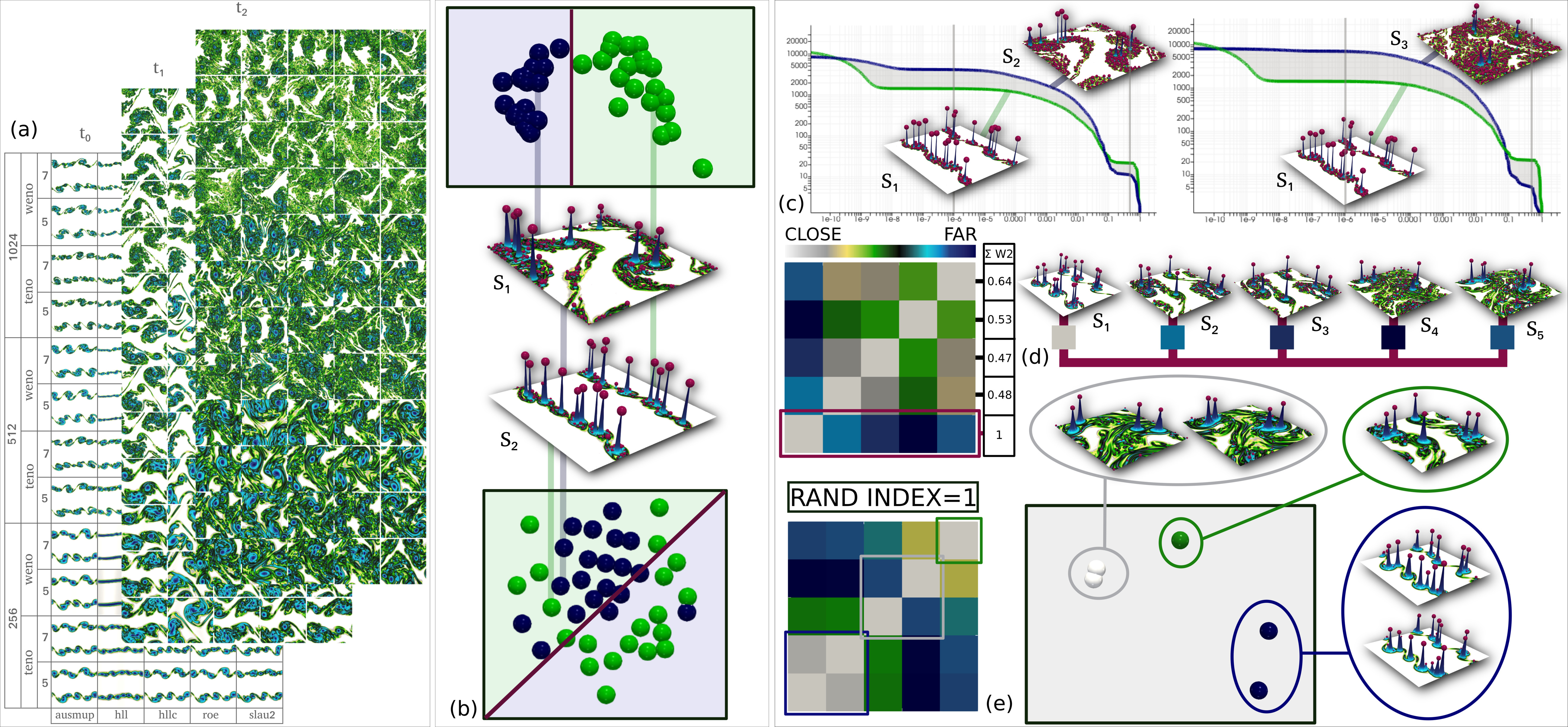}
  \caption{Topological Data Analysis protocols 
applied
  on an ensemble dataset of a Kelvin-Helmholtz instability.
(a) The 180 members of the ensemble obtained with variations of timesteps, interpolation schemes, orders, resolutions and Riemann solvers (\autoref{tab_parameters}).
(b) The top cluster represents the time separation of $t_0$ and $t_1$ for the 
flows $S_1$ and $S_2$ with the Wasserstein distance and the bottom cluster with 
the $L_2$-norm. Red lines show the timestep separation with our clustering 
method whereas the sphere colors are the ground truth, illustrating the 
limitation of the $L_2$-norm. 
(c) \textit{Persistence curve protocol}: Differences between integrals of persistence curves (gray area) of the enstrophy computed with a SLAU2 solver, an order 7 TENO scheme and a resolution of $1024\times 1024$ for various configurations ($S_1$ at $t_0$, $S_2$ and $S_3$ at $t_1$). These integral differences exhibit the appearance of vortices (critical points) as the time increases.
(d) \textit{Outlier distance protocol}: Wasserstein distance matrix for 5 configurations $S_1(t_0,HLLC)$, $S_2(t_1,Roe)$, $S_3(t_1,HLLC)$, $S_4(t_2,Roe)$, $S_5(t_2,HLLC)$ computed with an order 7 WENO-Z interpolation scheme at $512\times 512$.
The sum of each row
the configuration maximizing this distance between solvers and timesteps, here $S_1$.
(e) \textit{Unsupervised classification}: Wasserstein distance matrix for the 
previous configurations with an order 7 WENO-Z interpolation scheme at 
$256\times 256$. The clustering based on the Wasserstein distance and colored 
according to the Kmeans clustering method successfully segments the time 
steps.}
  \label{fig_teaser}
}

\usepackage{xcolor}
\usepackage{enumitem}

\newcommand{\domain}{\mathcal{M}}

\newcommand{\range}{\mathbb{R}}
\newcommand{\sublevelset}[1]{#1^{-1}_{-\infty}}

\newcommand{\Star}{St}
\newcommand{\Link}{Lk}
\newcommand{\simplex}{\sigma}
\newcommand{\face}{\tau}
\newcommand{\lowerlink}{\Link^{-}}
\newcommand{\upperlink}{\Link^{+}}
\newcommand{\Index}{\mathcal{I}}

\newcommand{\diagram}{\mathcal{D}}
\newcommand{\persistentCurve}{\mathcal{C}}
\newcommand{\wasserstein}[1]{W_#1}

\newcommand{\x}{\phantom{x}}

\newcommand{\isovalue}{w}

\newcommand{\cutout}[1]{\textcolor{blue}{#1}}
\renewcommand{\cutout}[1]{}

\newcommand{\eqSpace}{\vspace{-.5ex}}
\newcommand{\figureShrink}{0.9915}

\newcommand{\mycaption}[1]{
\vspace{-2ex}
\caption{#1}
\vspace{-2ex}
}

\begin{document}


\firstsection{Introduction}

\maketitle

Flow turbulence is a phenomenon of major importance in fluid dynamics.
It is characterized by chaotic changes in the motion of a flow (\emph{e.g.}
typical cigarette smoke patterns), which have a drastic impact in numerous
applications (aeronautics, weather forecast, climate modeling, material
sciences, astronomy, etc.).
Although turbulence has been studied since the early stages of modern physics,
its theoretical mastery remains incomplete \cite{baez2006} and the
understanding of the
Navier-Stokes equations, central
in the description of
fluid motion, is still
considered as
a
major open challenge in mathematics and physics,
as proven by the Clay Mathematics Institute selecting it to be
among its celebrated Millennium Prize problems \cite{fefferman2000}.
Thus, in engineering applications, the main practical solution available for
the study of turbulence
remains numerical simulation.

\noindent
\textbf{Problem introduction:}
However,
many
different numerical solvers can be used to simulate a given flow configuration, each
solver being itself subject to several input parameters (such as domain
resolution, interpolation scheme and order, etc.): when faced with such a wide variety,
the main problem for users becomes
the configuration itself of the simulation parameters.
In particular, domain experts want not only to identify
the solver configurations which produce the most realistic simulations, but
they also want to discover configurations resulting in
degraded but fast
computations, which still produce simulations of acceptable realism. The
fundamental problem behind such comparative analyses is that of comparing
\emph{quantitatively} turbulent flows. For instance, the quantitative realism
of a simulation could be evaluated by comparing its outcome to a reference,
either obtained by acquisition or by a highly detailed simulation considered as
a ground-truth. However, the chaotic nature of turbulent flows makes their direct
comparison with standard imaging tools
impractical. For instance, turbulent flows which are considered similar at a
high level by domain experts (in terms of the number and size of their vortices)
are reported by classical metrics such as the $L_2$ norm as being very distant
(\autoref{fig_teaser}), as such pointwise measures are sensitive to mild
geometrical variations in the data and miss global structural
similarities in such a chaotic context. This observation motivates the
consideration of alternative similarity estimation tools, which focus on the
\emph{structure} of the flow, rather than its raw geometry.

In that regard, Topological Data Analysis (TDA) \cite{edelsbrunner09} forms a
family of generic, robust and efficient techniques, whose purpose is precisely
to recover hidden implicit structural patterns in complex data, and to enable
their reliable representation and comparison. As such, they provide
potentially relevant alternatives to standard comparison measures used in
scientific imaging, such as the $L_2$ norm. Moreover, the utility of TDA has
been already demonstrated in a number of analysis and visualization tasks
\cite{heine16}, with examples of successful applications in
combustion \cite{laney_vis06, bremer_tvcg11, gyulassy_ev14},
material sciences \cite{gyulassy_vis07, gyulassy_vis15, favelier16},
bioimaging \cite{carr04, topoAngler, beiBrain18},
quantum chemistry \cite{chemistry_vis14, harshChemistry, Malgorzata19}, or
astrophysics \cite{sousbie11, shivashankar2016felix}. In
particular, the critical points of flow vorticity indicators have been
reported to appropriately capture the center of vortices \cite{kasten_tvcg11,
bridel_ldav19}, as well as their importance, with the notion of
\emph{topological persistence} \cite{edelsbrunner02}. Such  results provide
additional evidence and consolidate the intuition that TDA could be a
relevant framework for comparing turbulent flows.

The ambition of this application paper is to provide a comprehensive
experimental evaluation of
the above intuition, i.e. to assess the suitability of
topological data representations and their associated analysis tools for the
quantitative comparisons of turbulent flows.
We shall focus on a specific type of turbulence that expresses itself in two dimensions (Kelvin-Helmholtz instability): it is a fair representative of generic turbulence (\emph{a.k.a.} three-dimensional viscous turbulence) and allows for affordable high-resolution simulations to feed our study.
Specifically, our study documents
the usage of the persistence diagram of the maxima of flow enstrophy (an established indicator of vorticity for two-dimensional flows,
\autoref{sec_background}) for the topological representation of 180 members of
an ensemble of hydrodynamic turbulent flows, generated by a coarse sampling of
the parameter space of five distinct solvers.
We document five main hypotheses (\autoref{sec_caseStudy}) reported by domain
experts, describing their expectations with regard to the variability among the
flows generated by the distinct solver configurations.
Then, we describe three evaluation protocols (\autoref{sec_protocols}) designed
to assess the validation of the above hypotheses by standard comparison
measures ($L_2$ norm) on one hand, and by topological methods on the other.
Specifically, these protocols exploit the persistence curve, the
$L_2$-Wasserstein distance between persistence diagrams \cite{Turner2014} and
$k$-means in the $L_2$-Wasserstein metric space \cite{vidal_vis19}. Finally, we
document
the
results of these
protocols on the input ensemble
(\autoref{sec_experimentalResults}). We believe that the insights reported by
our study bring a strong experimental evidence of the suitability
of TDA for representing and comparing turbulent flows, thereby providing
confidence in its usage by the fluid dynamics community in the future.
Moreover, our flow data and evaluation protocols provide to the TDA community
an application-approved benchmark for the evaluation and design of further
topological distances in the future.

\begin{figure}
 \centering
 \includegraphics[width=\figureShrink\linewidth]{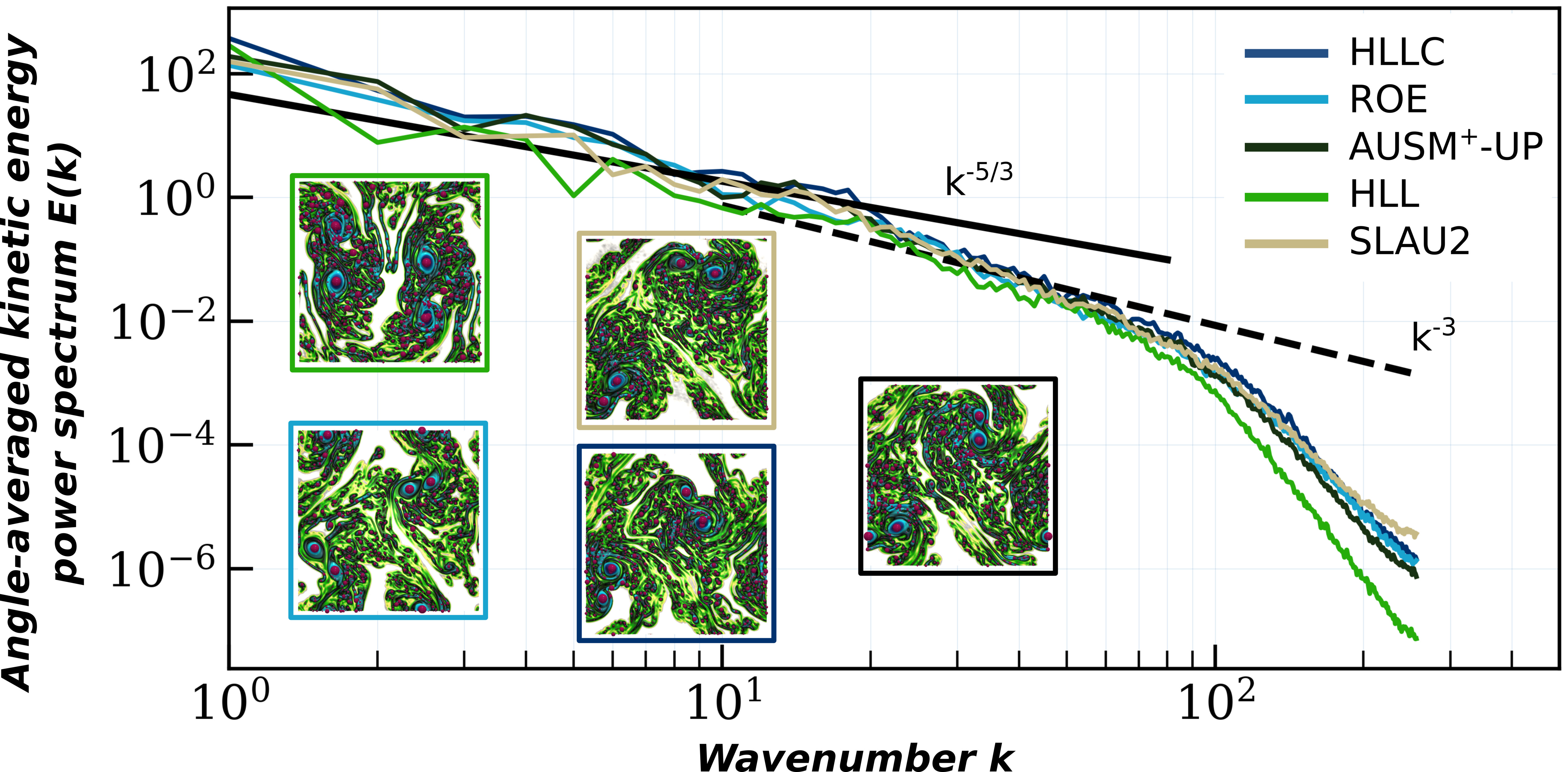}
 \mycaption{Baseline analysis by angle-averaged kinetic energy power spectrum
 for all solvers (WENO-Z, order $7$, $t_2$, $512\times512$).
 }
 \label{energie}
\end{figure}

\subsection{Related work}
\label{sec_relatedWork}
This section presents the literature related to our work.
First, we
 discuss previous work dealing with \emph{(i)} simulations of turbulent
flow
and their quantitative evaluation. Next, \emph{(ii)} we review
previous work dealing with topological methods for the analysis of ensemble
data.

\noindent
\textbf{\emph{(i)} Turbulent flow simulation.}
Turbulence is ubiquitous in nature, at all scales, from Higgs-Boson condensates \cite{Kraichnan1967} to a stirred cup of coffee to geophysical flows \cite{Kraichnan1980} to galaxy formation.
While a significant literature in graphics \cite{KimTJG08, ZhangLSSQ14, BaiWDL21} focused on the efficient generation of visually plausible turbulence, we focus in this work on the direct numerical simulation of the underlying physical equations, for engineering applications.
A special distinction of 2D turbulence is that it is never realized in nature that, unless strongly constrained,
always has some degree of three-dimensionality but rather it only exists in computer simulations.
Two-dimensional turbulence has thus been studied extensively by the latter means \emph{e.g.} for its importance as an idealization of meteorological flows \cite{Boffetta2011}, its role in the confinement of thermonuclear plasmas \cite{Kraichnan1980} but also as a cost-effective numerical testing ground for three-dimensional flows dynamical theories \cite{Tabeling2002}.
Most such studies focus either on validating predictions of theorists \cite{Kraichnan1967,lilly1989two} or on providing insights into the dynamic behavior of 2D eddies thanks to high-resolution simulations \cite{lilly1989two,Maltrud1991}.
The simulations are usually analyzed by considering macroscopic quantities such as the enstrophy (see below equation~\eqref{eq:enstrophy}), or by considering the Fourier decomposition of the 2D field (similar to \autoref{energie})- integral indicators that make it near impossible to compare and/or classify the results of, say, a parametric study.
Some efforts have been made recently \cite{san2015evaluation} to provide the users with some guidelines to best choose the numerical methods and parameters for the simulation of 2D turbulence, but still using, mostly, the aforementioned integral, inaccurate, indicators.
The present study aims at providing the workers in 2D turbulence with another tool to classify their results and best choose their settings, this time using mostly local indicators able to exploit the whole flow.

\noindent
\textbf{\emph{(ii)} Topological methods for ensemble analysis.}
Concepts and algorithms from computational topology \cite{edelsbrunner09} have
been investigated, adapted and extended by the visualization community for more
than twenty years \cite{heine16}. Specifically, a large body of literature has
been dedicated to the analysis and visualization of flow data with
topological methods and we refer the readers to a series of surveys on the topic
\cite{ScheuermannT05, GarthT06,
LarameeHZP07, PobitzerPFSKTMH11,
wang2016,
BujackYHGW20},
including a
recent iteration \cite{flowIntro21}.
A substantial line of work
\cite{petz, otto2, otto1} focused on extending topological techniques to
uncertain vector fields, where flow variability is encoded via a pointwise
estimator (\emph{e.g.} an histogram) of an \emph{a priori} vector distribution, but only few
techniques explicitly focused on the analysis of flow variability in an
ensemble.
Specifically, several comparative visualization techniques have been
proposed \cite{filip3, schneider2012,Hummel2013, GuoHPSCH16, JaremaKW16, filip2, filip1, LohfinkG20}. Ferstl et al.
\cite{Ferstl2016} investigated the global structure of flow ensembles, by
proposing a  clustering approach of the  members, based on a measure of
streamline similarity.
However, these techniques assume a mild
geometrical variability within the ensemble and are therefore
not suited to highly
turbulent flows as studied in this work, where the geometry of the features
(streamlines, vortices) chaotically changes from one ensemble
member to the other (\autoref{fig_teaser}), even upon only slight variations of
the simulation input parameters.
In certain application contexts, CFD experts often prefer to focus their analysis on (simpler to interpret) scalar descriptors generated from the flow, such as the kinetic energy (\autoref{energie}) or the enstrophy (\autoref{eq:enstrophy}). Such a transition to a scalar descriptor enables them to leverage the existing tools for scalar data analysis.
For instance, several
authors \cite{kasten_tvcg11, bridel_ldav19} have shown that, given a relevant
vorticity scalar descriptor, the center of the flow vortices could be reliably
extracted and tracked over time, which supports the idea that
topological methods for scalar data can be useful for the description of
specific flow features. Thus, in the following, we describe the literature
related to topological methods for scalar data.
Popular topological representations include the persistence diagram
\cite{edelsbrunner02, edelsbrunner09}
which represents the
population of features of interest in function of their salience, and which
can be computed via matrix reduction
\cite{edelsbrunner09, dipha}. The Reeb graph \cite{biasotti08}, which describes
the connectivity evolution of level sets, has also been widely studied and
several efficient algorithms have been documented \cite{pascucci07,
tierny_vis09, parsa12, DoraiswamyN13}, including parallel algorithms
\cite{gueunet_egpgv19}.
Efficient algorithms have also been documented for its variants,
the
merge and contour trees \cite{tarasov98, carr00}
and parallel algorithms have
also been described \cite{MaadasamyDN12, AcharyaN15,
CarrWSA16, gueunet_tpds19}. The Morse-Smale complex \cite{EdelsbrunnerHZ01,
EdelsbrunnerHNP03,
BremerEHP03}, which depicts the
global behavior of integral lines, is another popular topological data
abstraction in visualization \cite{Defl15}. Robust and efficient algorithms
have been introduced for its computation  \cite{robins_pami11, ShivashankarN12,
gyulassy_vis18} based on Discrete Morse Theory \cite{forman98}.
Inspired by the literature in
optimal transport \cite{Kantorovich, monge81}, the Wasserstein distance between
persistence diagrams \cite{edelsbrunner09}
(and its variant the
Bottleneck distance \cite{edelsbrunner02}) have been extensively studied.
In
practice, it enables users to compare ensemble members based on their
persistence diagrams.
In our experimental study, we focus on this first, established topological tool
for comparing the topology of ensemble members and we refer the reader to a
recent survey \cite{YanMSRNHW21} for a description of alternative metrics
between topological descriptors.
Several techniques have been proposed for summarizing the topological features
in an ensemble or analyzing their variability.
Favelier et al.
\cite{favelier2018} and Athawale et al. \cite{athawale_tvcg19} introduced
approaches for analyzing the variability of critical points and
gradient separatrices respectively. Recent approaches aimed at summarizing
an ensemble of topological descriptors by computing a notion of \emph{average}
descriptor,
given a specific metric.
This notion has
been well studied for persistence diagrams \cite{Turner2014,
lacombe2018, vidal_vis19}, with direct applications to ensemble clustering
\cite{vidal_vis19}, and analogies have been developed for merge
trees \cite{YanWMGW20, pont_vis21}.
%
%
%
%


\subsection{Contributions}
This application paper makes the following new contributions:
\vspace{-1ex}
\begin{enumerate}[itemsep=-0.75ex]
 \item{\emph{An evaluation procedure for distances between turbulent
flows:}
 We document 5 main hypotheses reported by domain experts, describing their
expectations
about
flow variability
within the studied ensemble (180 members).
We contribute 3 evaluation protocols,
to assess
the validation of the
hypotheses by a specific metric,
given as a distance matrix between the members;}
 \item{\emph{A comprehensive experimental study:}
 We provide a detailed experimental study of the ensemble under consideration
for two distance metrics: \emph{(i)} a standard distance used in scientific
imaging (the $L_2$ norm) and \emph{(ii)} an established topological distance
between persistence diagrams (the $L_2$-Wasserstein metric). Our
experiments demonstrate the superiority of the topological distance to report
as close solver configurations which are expected to be similar by domain
experts. The insights reported by our study bring an experimental evidence
of the suitability of the persistence diagram for representing and comparing
turbulent flows, thereby providing to the fluid dynamics community confidence
for its usage in future work;}
 \item{\emph{An application-approved benchmark:}
 We provide as supplemental material \emph{(i)} the input ensemble of 180
turbulent flows (see \cite{data} for a direct link)
and
\emph{(ii)} a python template script for reproducing our results.
Our script supports the usage of custom distance matrices, thereby providing to 
the TDA community an application-approved template for the evaluation and design 
of further topological distances.
}

\end{enumerate}

\section{Background}
\label{sec_background}
This section presents the background used in this study in $(i)$ numerical simulation by presenting the equations, the interpolation schemes and the solvers implemented in our simulation code. Then the background in $(ii)$ topological data analysis introduces the main notions used such as critical points, persistence diagrams or the Wasserstein distant metric.

\subsection{Numerical simulation}
\label{sec_simulation}
\label{sec_solvers}

In this work, we consider the two-dimensional compressible unsteady Euler equations for inviscid flows \cite{masatsuka2013}:
\eqSpace
\begin{equation}
    \mathbf{U}_t + \mathbf{F}_x + \mathbf{G}_y = 0,
    \label{eq:cons_euler}
    \eqSpace
\end{equation}
\noindent where the subscripts indicate differentiation, $\mathbf{U}$ is the 
vector of conservative dimensionless variables and $\mathbf{F}$ and $\mathbf{G}$ 
represent the inviscid fluxes in $x$ and $y$ direction respectively.
Those vectors are defined as:
\eqSpace
\eqSpace
\eqSpace
\begin{equation}
    \begin{array}{l}
        \mathbf{U} = \left[\begin{array}{c}\rho \\ \rho u \\ \rho v \\ \rho E\end{array}\right], ~~
        \mathbf{F} = \left[\begin{array}{c}\rho u \\ \rho u^2 + p\\ \rho u v\\ (\rho E + p)u\end{array}\right], ~~
        \mathbf{G} = \left[\begin{array}{c}\rho v \\ \rho u v \\ \rho v^2 + p \\ (\rho E + p)v\end{array}\right].
    \end{array}
    \label{eq:cons_euler_vectors}
    \eqSpace
\end{equation}
\noindent In the above expressions, $t$ denotes the time and $x$ and $y$ are the Cartesian coordinates.
$\rho$ denotes density, $u$ and $v$ denote the $x-$ and $y-$
coordinates of the velocity vector $\mathbf{w}$,
$E$ denotes the specific total energy and $p$ denotes the static pressure.
The aforementioned mathematical model is described as it is implemented in the in-house code HYPERION (HYPERsonic vehicle design with Immersed bOuNdaries) whose primary capabilities as a massively parallel structured solver using immersed boundary conditions have already been discussed by Bridel-Bertomeu \cite{bridel2021immersed}.
The present study uses only regular Cartesian grids with constant grid spacings (grid of pixels) in both directions of space, $\Delta x$ and $\Delta y$, and will not rely on any immersed boundary condition during the computations presented later.
This being said, the finite-volume method \cite{leveque2002finite,trangenstein2007numerical,toro2013riemann} is then employed for space discretization of the compressible Euler equations~\eqref{eq:cons_euler}.

The 2D turbulence investigated in our work
is generated using a Kelvin-Helmholtz instability (see \cite{san2015evaluation} for a complete description)
simulated with high-order low-dissipation  reconstruction schemes of $5^{\text{th}}$- and $7^{\text{th}}$-order (\autoref{tab_parameters}).
The numerical fluxes between the cells are obtained using a variety of Riemann solvers detail at the end of this section.
To emulate turbulence in a infinite medium, all boundary conditions are set as periodic.
One common measure of turbulence in two dimensions that we will rely on is the 
local enstrophy $\mathcal{E}$, defined locally as the square of the flow vorticity:
\eqSpace
\begin{equation}
    \mathcal{E} = 0.5\left\vert\nabla\times\mathbf{w}\right\vert^2.
    \label{eq:enstrophy}
    \eqSpace
\end{equation}

When solving numerically the Euler equations  (\autoref{eq:cons_euler}), we 
start by interpolating the values of the flows at the cell interfaces.
Then, we have
to use an approximate Riemann solver
\cite{toro2013riemann}
to solve the eponymous problem 
on those interfaces.
In the remainder,
we will expose the different
methods used to make these two
calculations.

\noindent
\textbf{Interpolation schemes.} 
\label{sec_interpolations}
One problem in numerically solving the schemes is to be able to capture the strong discontinuities while capturing the small scales of the turbulence. In addition, we want to be as accurate as possible in our interpolation. To do this, researchers and engineers have developed several high-order reconstruction methods. A common scheme for solving compressible flows in the presence of strong discontinuities is the \textit{Weighted Essentially Non-Oscillatory (WENO)} scheme \cite{liu1994weighted}. Several variants of this scheme have been
introduced,
to improve its performances\cite{jiang1996efficient}\cite{hu2010adaptive}\cite{henrick2005mapped}.
We are particularly interested in two families: the well-known robust but dissipative WENO-Z \cite{borges2008improved} and the TENO (T for \emph{Targeted}) \cite{fu2016family}, which better discriminates scales \cite{hu2011scale}.

\noindent
\textbf{Solvers.}
We have to solve the Riemann problems at the interfaces between the cells of the mesh.
One solution is to use an exact Godunov solver \cite{toro2013riemann} which takes into account a large number of nonlinear operations - too expensive however when calculating complex flows. 
Rather, researchers and engineers are interested in approximate Riemann solvers. The most used approximate solvers can be grouped in three large families: Flux Difference Splitting (FDS), Flux Vector Splitting (FVS) and Flux Type Splitting (FTS) \cite{toro2013riemann}.
In this section we will focus on two types of solvers in particular, Flux Difference Splitting solvers that work as a finite volume method to solve the Riemann problem and Flux Splitting Riemann solvers that combine the qualities of the other two families by separating kinematic and acoustic scales.

\noindent
\textbf{HLL} \textit{(Harten, Lax, and van Leer)} : FDS
scheme developed by
Harten et al. \cite{harten1983upstream}.
It does not take into account
contact discontinuities, i.e. lines crossing two states.
For turbulent phenomena,
the interface between vortices will
therefore be less described.

\noindent
\textbf{Roe}  and \textbf{HLLC} \textit{(Harten, Lax, and van Leer with Contact)} : FDS type schemes developed by \cite{toro1994restoration}\cite{Roe1981approximate}. These two schemes are robust and thus allow to reproduce the strong discontinuity (shock) and takes into account the discontinuities of contact. Thus, with these schemes, the reconstruction of the vortices represented in our flows, is perform with more accuracy than with the HLL solver.

\noindent
In its study on low speed Riemann solvers, \cite{qu2014study} notice that the solvers are unable to obtain physical solutions. Therefore, there is a need for approximate Riemann solvers to accurately reconstruct both low and high speed flows. This is why we are interested in two flux type splitting (FTS) solvers, which take into account all velocities to obtain low Mach and high Mach physics solutions.

\noindent
\textbf{AUSM$^+$-UP} \textit{(Advection Upstream Splitting Method +UP)} : By adding improvements \cite{liou2006sequel} to the AUSM+ \cite{liou1996sequel} solver Liou increases its level of accuracy for all speeds. This new solver takes into account contact discontinuities, reconstructs also strong discontinuities and gives physical solutions for all speeds.

\noindent
\textbf{SLAU2} \textit{(Simple Low-Dissipation AUSM 2)} :
the analysis of the dissipation pressure term of the AUSM+ \cite{shima2009new} shows that it is too high for low speeds.  The author decided to control the pressure flux and implemented the SLAU solver.
This
has been extended
\cite{kitamura2013towards}
so that the dissipation becomes proportional to the Mach number. This solver takes into account contact discontinuities, reconstructs also strong discontinuities and gives physical solutions for all speeds.

\begin{figure}
 \centering
 \vspace{-1ex}
 \includegraphics[width=\figureShrink\linewidth]{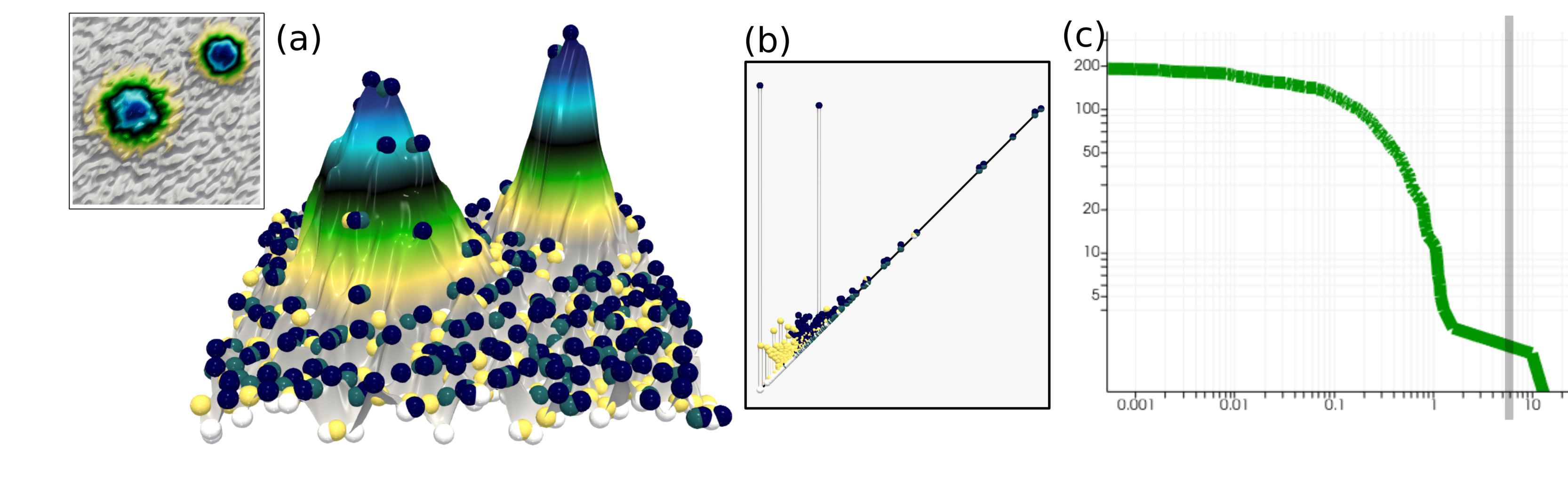}
 \mycaption{Critical points (spheres, white: minima, blue: maxima, other:
saddles), persistence diagram (b), persistence curve (c) of
a noisy (a) 2D
scalar field. The persistence diagram captures the main two hills of the
terrain as prominent persistence pairs (large vertical segments), while small
oscillations due to noise induce features near the diagonal.}
 \label{fig_toyTopology}
\end{figure}

\subsection{Topological data analysis}
\label{sec_topology}

This section presents the topological background of our work. It contains
definitions adapted from the Topology ToolKit \cite{ttk17}.
We refer the
reader to textbooks \cite{edelsbrunner09} for an introduction to
topology.

\noindent
\textbf{Input data.}
The input data is given as an ensemble of $N$ piecewise linear (PL) scalar
fields
$f_i : \domain \rightarrow \range$, with $i \in \{1, \dots,  N\}$, defined on a
PL 2-manifold $\domain$.
Specifically, $f_i$ represents the pointwise flow enstrophy
(\autoref{sec_simulation}) and $\domain$ is the Freudenthal triangulation
\cite{freudenthal42, kuhn60} of a $2$-dimensional regular grid, which is
periodic in both dimensions ($\domain$ is homeomorphic to a $2$-dimensional
torus). The triangulation is performed implicitly, by emulating the
simplicial structure upon traversal queries. Thus it induces no
memory overhead \cite{ttk17}.
The scalar values are given at the vertices of $\domain$ and are linearly
interpolated
on the simplices of higher dimensions.
$f$ is assumed to be injective on the vertices  of $\domain$.
This is
enforced in practice with a symbolic
perturbation inspired by Simulation of Simplicity \cite{edelsbrunner90}.

\noindent
\textbf{Critical points.}
Topological features in $f_i$ can be tracked with the notion of
\emph{sub-level set}, noted
$\sublevelset{{f_i}}(\isovalue)=\{p \in \domain~|
~f_i(p) < \isovalue\}$. It is defined as the pre-image of  $(-\infty,
\isovalue)$
by
$f_i$.
In particular, the topology of these sub-level sets (in 2D, their connected
components and cycles) can only change at special locations.
As $\isovalue$ continuously increases, the topology of
$\sublevelset{{f_i}}(\isovalue)$ changes at specific vertices of $\domain$,
called the \emph{critical points} of $f_i$ \cite{banchoff70}, defined next.
The \emph{star} of a vertex $v \in \domain$, noted $\Star(v)$,  is
the set of its co-faces:
$\Star(v) = \{ \simplex \in \domain ~|~ v < \sigma \}$.
It can be interpreted as the smallest combinatorial neighborhood around $v$.
The \emph{link} of $v$,
noted $\Link(v)$, is the set of the faces $\face$ of the simplices $\simplex$
of $\Star(v)$ with empty intersection with $v$:
$\Link(v) = \{ \face \in \domain ~ | ~ \face < \simplex, ~
\simplex\in \Star(v), ~ \face \cap v = \emptyset\}$.
The link of a vertex can
be interpreted as the boundary of its star.
The \emph{lower link} of $v$, noted $\lowerlink(v)$, is given by the
set of simplices of $\Link(v)$ which only contain vertices \emph{lower} than
$v$:
$\lowerlink(v) = \{ \simplex \in \Link(v) ~ | ~ \forall v' \in \sigma, ~ f_i(v')
< f_i(v)\}$. The upper link is defined symmetrically: $\upperlink(v) = \{
\simplex \in \Link(v) ~ | ~
\forall v' \in \sigma, ~ f_i(v') > f_i(v)\}$.
A vertex $v$ is \emph{regular}  if
and only if
both $\lowerlink(v)$ and $\upperlink(v)$ are simply connected. For
such vertices, the sub-level sets do not change their topology as they span
$\Star(v)$. Otherwise, $v$ is
a \emph{critical point}.
These can be classified with regard to their
\emph{index} $\Index(v)$.
It is equal to $0$ for local minima
($\lowerlink(v) = \emptyset$), to 2 for local maxima
($\upperlink(v) = \emptyset$) and otherwise to 1 for
saddles (\autoref{fig_toyTopology}a).
In practice, $f_i$ is enforced to contain only isolated, non-degenerate
critical points \cite{edelsbrunner90, edelsbrunner03}.
In the case of
the pointwise flow enstrophy, local maxima denote the center of vortices in the
turbulent flow. However, since the critical point characterization is based on a
classification which is only local (restricted to the link of each vertex), the
slightest oscillation in the data results in practice in the appearance of
spurious critical points, especially in the case of noisy data such as
turbulent flows. This motivates the introduction of an importance measure on
critical points, discussed next, in order to disambiguate vortices from noise.

\noindent
\textbf{Persistence diagrams.}
Several important
measures for critical points have been studied \cite{carr04},
including \emph{topological persistence} \cite{edelsbrunner02}, which is tightly
coupled to the notion of Persistence diagram \cite{edelsbrunner09}, which we
briefly describe here.
Persistence
assesses the importance of a critical point, based on the
lifetime of the topological feature it created (or destroyed) in
 $\sublevelset{{f_i}}(w)$, as one continuously increase the isovalue $w$.
In particular, as $w$ increases, new connected components of
$\sublevelset{{f_i}}(w)$ are created at the minima of $f_i$. The Elder rule
\cite{edelsbrunner09} indicates that if two connected
components, created at the minima $m_0$ and $m_1$ with $f_i(m_0) < f_i(m_1)$,
meet
at a given saddle $s$, the \emph{youngest} of the two components (the
one created at $m_1$) \emph{dies} in favor of the \emph{oldest} one (created at
$m_0$). In this case, a \emph{persistence pair} $(m_1, s)$ is created and
its
\emph{topological persistence} $p$ is given by $p(m_1, s) = f_i(s) - f_i(m_1)$.
All the local minima
can be
unambiguously
paired following this strategy, while the
global minimum is usually paired, by convention, with the global maximum.
The symmetric reasoning
can be applied to characterize, with saddle/maximum pairs, the life
time of the independent cycles of
$\sublevelset{{f_i}}(w)$.
Persistence pairs are usually
visualized with the \emph{Persistence diagram} $\diagram(f_i)$
\cite{edelsbrunner09}, which embeds each pair $(c, c')$, with $f_i(c) <
f_i(c')$,
as a point in the 2D plane, at location $\big(f_i(c), f_i(c')\big)$.
The  value $f_i(c)$ is 
called the \emph{birth} of the
feature, while $f_i(c')$ is called its \emph{death}.
The
pair
persistence
can be
visualized as the height of the point to the diagonal.
Features with a high persistence stand out, away from the diagonal,
while noisy features are typically located in its vicinity 
(\autoref{fig_toyTopology}b).
The conciseness, stability \cite{edelsbrunner02} and expressiveness of this
diagram made it a popular tool
for data summarization tasks, as
it provides visual hints about the number, ranges and salience
of the features of interest.
To compare two datasets $f_i$ and $f_j$, persistence diagrams can be efficiently compared with the notion of $L_2$-Wasserstein distance \cite{CohenSteinerEH05,
Turner2014, Kerber2016} (we leave the practical study of distances between more
advanced topological descriptors  \cite{SridharamurthyM20,
pont_vis21} to future work). This distance is based on a bipartite assignment optimization problem (between the points of the two diagrams to compare), for which exact \cite{Munkres1957} and approximate \cite{Bertsekas81}
implementations are publicly available \cite{ttk17, ttk19}.
Specifically, we use in our approach the fast approximation scheme by Vidal et al. \cite{vidal_vis19}.
We refer the reader to \cite{Kerber2016, vidal_vis19, ttk19} for further
details.
Once the $L_2$-Wasserstein distance between two diagrams $\diagram(f_i)$ and $\diagram(f_j)$ is available (noted $\wasserstein{2}\big(\diagram(f_i), \diagram(f_j)\big)$), more advanced geometrical objects can be considered, such as \emph{Wasserstein barycenters} \cite{Turner2014, vidal_vis19}, which are
diagrams minimizing the sum of their distance to an ensemble of diagrams, and which consequently, can be considered as a reliable representative of the ensemble.
This notion of \emph{barycenter}
is conducive to the design of clustering algorithms.
The $k$-means algorithm
can be easily extended,
by using $\wasserstein{2}$ to measure distances between diagrams, and by
considering as cluster centroid, at each iteration of the $k$-means, the
 barycenter of the cluster.

\noindent
\textbf{Persistence curves.}
A popular, alternate, representation of persistence features is the notion of
\emph{Persistence Curve}, noted $\persistentCurve(f_i)$, which plots the
population of persistent pairs as a function of their persistence.
Specifically, it encodes the number of pairs (Y axis) whose persistence is
\emph{larger} than a threshold $\epsilon$ (X axis).
For $X=0$, $Y$ is equal to the total number of persistence pairs, while for the
largest values of $X$, $Y$ indicates the number of prominent, high-persistence
features (\autoref{fig_toyTopology}c).
In practice, large
plateaus in this curve will indicate \emph{stable} persistence ranges, for
which no (or few) topological features are present in the data. These 
correspond to \emph{separations} (vertical line, \autoref{fig_toyTopology}c) 
between populations of topological
features of distinct persistence scales, typically the noise (low $X$
values) and the persistent features (high $X$ values).

\section{Case Study}
\label{sec_caseStudy}
In this section, we give \emph{(i)} a description of the ensemble (made publicly
available \cite{data})
representing the Kelvin Helmholtz Instabilities (KHI) computed on 
our institution's
facilities.
Next, we state \emph{(ii)} the challenges in understanding
such phenomena and we
provide
\emph{(iii)} theoretical hypothesis that our
experimental protocols have to verify.   

\begin{figure}
 \centering
 \vspace{-1ex}
 \includegraphics[width=\figureShrink\linewidth]{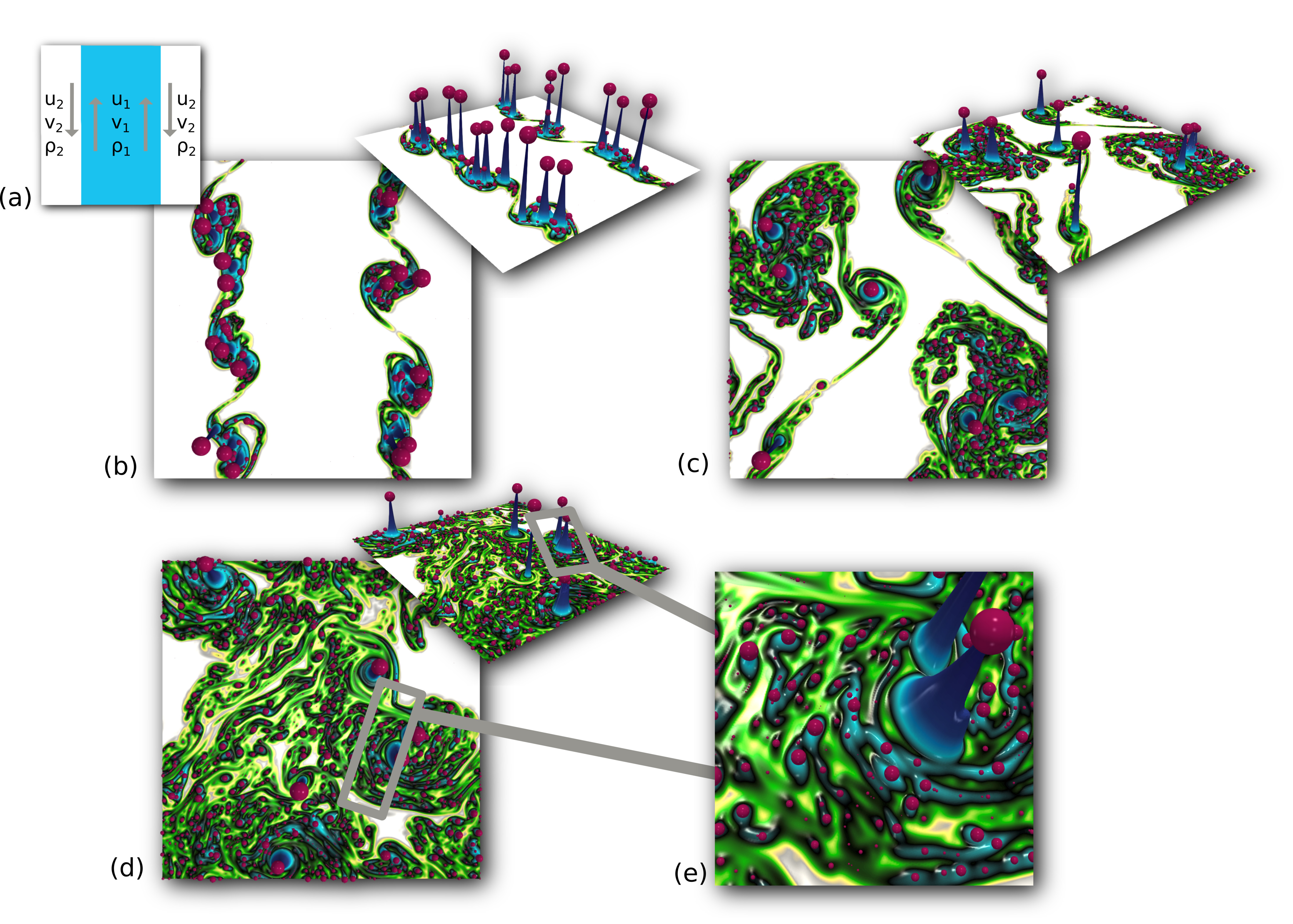}
 \vspace{-2ex}
 \mycaption{Initialization of the Kelvin-Helmholtz instability (a). This simulation was obtained
with the AUSM$^+$-UP solver with a TENO 5 order interpolation at physical times $0.25$(b), 
$0.75$(c) and $1.25$(d). Red spheres scaled by the persistence represent the maximum critical points. Zoom of the turbulence structures (e).}
  \vspace{2ex}
 \label{initcase}
\end{figure}

\subsection{Data description}
The initialization of the KHI was generated with two fluids of different 
densities ($\rho_{1}$, $\rho_{2}$) (\autoref{initcase}a). The
different velocities of opposite direction ($\{u _{1},v_{1}\},\{u _{2},v_{2}\}$) 
of the fluids create a shearing zone where the turbulence appears with the KHI 
(\autoref{initcase}b). While the instability develops over time the main 
vortices grow (\autoref{initcase}c). After a longer simulation time, the main 
structures keep evolving (\autoref{initcase}d) and a large number of 
small-scale vortices appear in the vicinity of large-scale vortices
(\autoref{initcase}e) 
leading to a complex turbulent flow. This variation in vortex scale,
in addition to the chaotic flow geometry,
is notoriously
challenging for the analysis of turbulent flows.

\begin{table}[]
\centering
\scalebox{0.8}{
\begin{tabular}{|c|c|c|c|c|c|c|}
\hline
Parameter            & Resolution           & Order                &  Time       
       &  Solver              &  Scheme              &  Total              \\ 
\hline
\hline
                     &                      &                      &             
         &  HLL            &                 &                \\
                     &  256                 &      5               &       t$_0$ 
         &  SLAU2          &  TENO         &                \\
Value                &  512                 &      7               &       t$_1$ 
         &  AUSM$^+$-UP         &  WENO-Z        &                \\
                     &  1024                &                      &       t$_2$ 
         &  Roe            &                &                \\
                     &                      &                      &             
         &  HLLC           &                &                \\                  
                                             
\hline
\hline
Number&3&2&3&5&2&180\\
\hline
\multicolumn{1}{l}{} & \multicolumn{1}{l}{} & \multicolumn{1}{l}{} & 
\multicolumn{1}{l}{} & \multicolumn{1}{l}{} & \multicolumn{1}{l}{} & 
\multicolumn{1}{l}{} 
\end{tabular}
}

\caption{Parameter space of the HYPERION simulation code leading to a total of 180 members for the ensemble dataset used in this study.}
\vspace{-3ex}
\label{tab_parameters}
\end{table}


The HYPERION simulation code introduced in \autoref{sec_simulation} has been 
used to generate the ensemble dataset. All the simulations have been run on a 
supercomputer 
at our institution. Each simulation have been executed in parallel using 16 MPI processes and have been distributed over the supercomputer. The total 
simulation took about 745 CPU hours. The raw data has been dump on disk with the 
metadata stored in XDMF files and the scalar fields in HDF5 files leading to 14 
GB for the entire ensemble dataset. We processed these results to extract the 
enstrophy scalar field (\autoref{eq:enstrophy}) and stored it to a VTK file 
format \cite{Kitware:2003} using an image data structure for regular grids 
(VTI). This reduces the entire ensemble to 600 MB. 

The ensemble dataset corresponds to different computational configurations for 
the same turbulent instability. HYPERION handles different parameter types such 
as scalars or enumerations, which allows the users to compute various numerical 
simulations in the same parametric study. The resolution of the 2D regular grid, 
the simulation time, the interpolation scheme, the order of interpolation and 
the Riemann solvers presented in \autoref{sec_background} are our different 
parameters. \autoref{tab_parameters} details the parameter types and values as 
well as the number of samples per parameter, leading overall to an ensemble of 
$3\times2\times3\times5\times2=180$ members illustrated \autoref{fig_teaser}a. 
Each parameter value of \autoref{tab_parameters} used to run the simulation has 
been stored as meta-data in the VTI files (i.e. \emph{Field Data} in the VTK 
terminology)  to keep track of the computational configuration for later 
analysis down the pipeline.
In order to ease the exploration of the ensemble dataset, we defined a 
SQL-type database using the cinema database feature of TTK \cite{ttk17, ttk19}. 
This representation facilitates the extraction of sub-samples of the ensemble, 
based on standard SQL queries on the simulation parameters 
(\autoref{tab_parameters}).


\begin{figure}
 \centering
 \vspace{-1ex}
 \includegraphics[width=\figureShrink\linewidth]{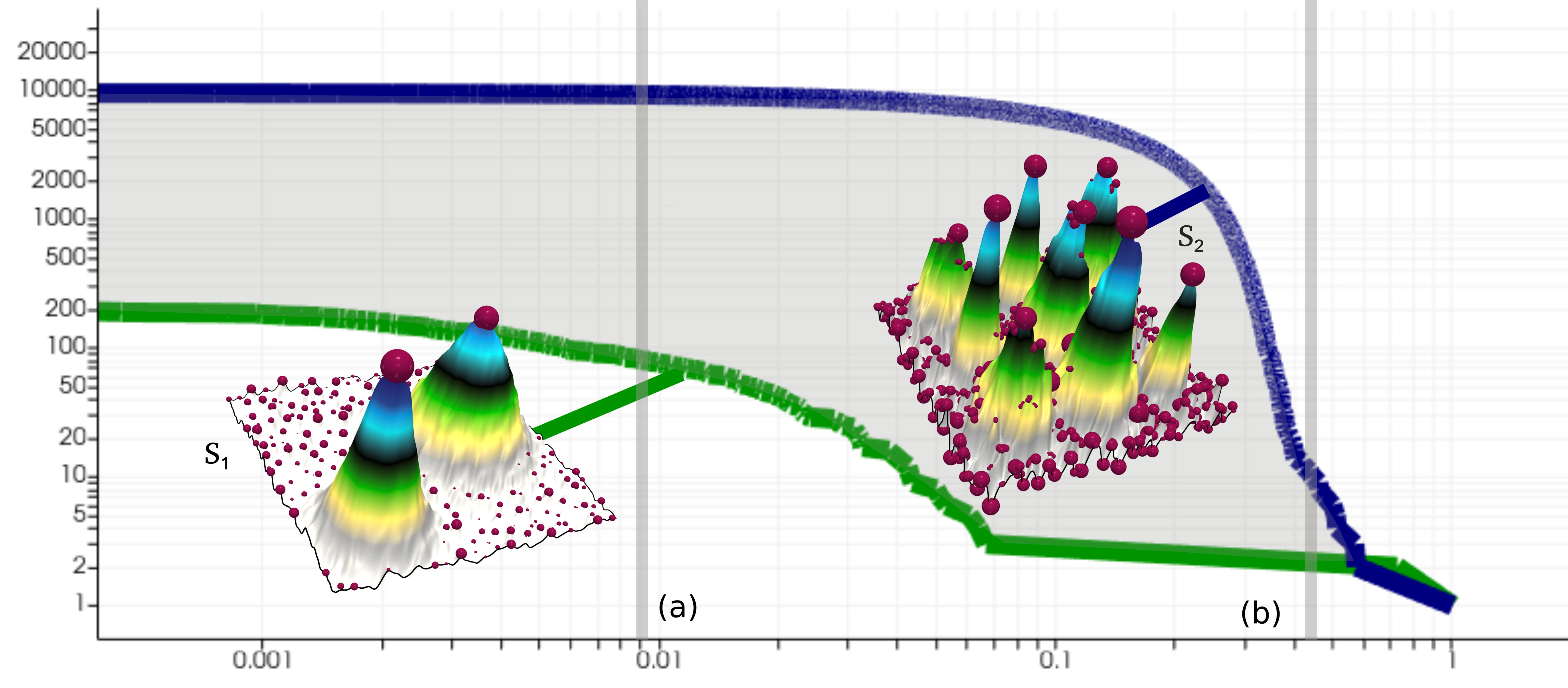}
 \mycaption{Persistence curves for two input scalar fields (X-axis: persistence threshold, Y-axis: number of maxima more persistent than X). $S_1$ generated with two Gaussian functions with noise and $S_2$ with 10 Gaussian functions with a stronger noise. Maxima critical points are represented by red spheres scaled by persistence. Vertical line corresponds to
 (a) small persistence critical points, (b) to high persistence. The grey area is the integral difference between the two curves.}
 \label{fig_gaussian_curve}
\end{figure}

\subsection{Problem statement}

Vehicle design, be it in an automotive or aeronautical context, is well known for its high number of constraints that are nowadays most often handled with the help of computational techniques.
In the aeronautical world for instance, engineers today face an incredible challenge wherein they have to be able to predict, at the same time, integral quantities at the wall of the vehicle such as heat flux or pressure as well as three-dimensional phenomena such as flow discontinuities and turbulence.
In other words, engineers have to deal with multiple types of physics and phenomena that have markedly different length- and time-scales but whose interactions are still of great importance to the accuracy of their predictions.
With limited time and resources to conduct the computer-aided simulations, the traditional approach is to rely on numerical strategies that temporally average most of the three-dimensional phenomena and rely more or less on models of turbulence to yield a fast and reasonable forecast.

Even in such a context of approximate simulations, the choice of the ingredients of the numerical recipe matters - methods of reconstruction, Riemann solvers, etc.
Making the right choices can indeed bring a significant increase in fidelity to the engineer, especially in terms of turbulence, by lessening the need for modeling and henceforth bring more margin in the design of the vehicle.
Turbulence is however by nature a chaotic phenomenon and conducting a systematical study of the impact of the different
numerical ingredients thereupon might prove tricky for a simple reason: beyond a certain level of accuracy, everything will \emph{look} the same.
Detecting the benefits of one method compared to another in that situation will be next to impossible - that is, with traditional techniques.
We propose here to use the ability of topological analysis to discern features that stay otherwise hidden in traditional fluid dynamics postprocessing to help with the choice of the right numerical ingredients.

\subsection{CFD Hypotheses}
\label{sec_hypotheses}
This section introduces the hypotheses provided by CFD experts, documenting their expectations about ensemble flow variability.


\label{Hypotheses}
\noindent
\textbf{Hypothesis H1.}
TENO induces more turbulence (i.e. more critical points) than
WENO-Z, for all configurations.\\
\textbf{Hypothesis H2.}
Order 5 and 7 are equivalent for Kelvin Helmholtz instabilities.\\
\textbf{Hypothesis H3.}
The HLL solver should provide a significantly distinct description, for all configurations.\\
\textbf{Hypothesis H4.}
The HLLC and Roe solvers should provide equivalent outputs for all configurations.\\
\textbf{Hypothesis H5.}
The
SLAU2 and AUSM$^+$-UP solvers should provide equivalent outputs for all configurations.\\

The above hypotheses are direct consequences of
observations, or
design choices. For instance, the TENO scheme has been reported to capture turbulence more accurately \cite{peng2021efficient}, which is expressed by Hypothesis H1. Similar kinetic energy curves (\autoref{energie}) have been reported for the orders 5 and 7, which is expressed by Hypothesis H2. The HLL solver, which is a dissipative approach, is known to model contact discontinuities poorly in contrast to more recent solvers, which is expressed in Hypothesis H3 \cite{toro2013riemann}. Finally, unlike the SLAU2 and AUSMUP (FTS type) solvers, the HLLC and RoE (FDS type) solvers have been reported to provide unphysical results at both low and high velocities (resulting in local oscillations in pressure and density), which is expressed in Hypotheses H4 and H5. 

From a practical point of view, the validation of these hypotheses has a major impact for the engineers when setting up their simulations. For instance, the validation of the Hypothesis H1 would justify the usage of a more computationally expensive scheme (TENO), while the validation of the Hypothesis H2 would enable the usage of less computationally expensive orders (5 instead of 7). Finally, the validation of the Hypotheses H3, H4, and H5 would help engineers properly select the most appropriate solvers, based on their flow characteristics.
Then, overall, the validation of these hypotheses
would provide reliable rules-of-thumb for the tuning of the solvers,
to achieve the best balance between accuracy and speed.


\subsection{Baseline analysis}
Traditional approaches for turbulent data analysis (\autoref{energie}) are based on an average of
quantities of interest, such as flow energy (\autoref{sec_relatedWork}).
The $L_2$ norm is another established distance for comparing scalar fields. Both strategies bear similarities in their averaging artifacts: they cannot distinguish the contribution
of small structures from the global flow, because these are masked by the weight
of larger vortices.
%
Moreover, the $L_2$-norm is also very sensitive to mild geometric variations,
whereas the chaotic nature of turbulent flows induces major geometric
variations between ensemble members.
This motivates the usage of  topological methods to capture
features in the KHI that will help us compare the members
(\autoref{sec_experimentalResults}). In the remainder, we will systematically compare our protocols based on topological distances (\autoref{sec_protocols}) to the $L_2$ norm, considered as the baseline approach, and detailed comparisons will be provided (\autoref{sec_experimentalResults}).

\section{Evaluation protocols}
\label{sec_protocols}
In this section, we present 3 protocols which can be used to verify the
hypotheses detailed in \autoref{Hypotheses}. One can directly use these
algorithms on the ensemble dataset. It corresponds to $(i)$ the separation of
the schemes and the independence of the orders, $(ii)$ the unique behavior of
the HLL solver and $(iii)$ similarities in class of solvers.

\subsection{Persistence curves}
\label{sec_persistence}
With this protocol (illustrated on toy examples, \autoref{fig_gaussian_curve}), we want to validate hypothesis H1 (\autoref{Hypotheses}) to discriminate the interpolation schemes TENO and WENO-Z regarding the differences in the enstrophy field. With this protocol, we also want to validate hypothesis H2 (\autoref{Hypotheses}) to confirm the independence of the orders \cite{san2015evaluation}. To better characterize the vortices influencing the turbulence, we use persistence curves (\autoref{sec_topology}). These curves will
allow us to threshold the structures (the eddies) at different scales and
thus to easily compare the number of small (\autoref{fig_gaussian_curve}a) and large (\autoref{fig_gaussian_curve}b) eddies using the integral of the persistence curve.

For the differentiation of the schemes, we take 5 simulation configurations where the physical time ($t_0,t_1,t_2$), the resolution ($256\times 256$, $512\times 512$, $ 1024\times 1024$) and the order (5,7) are fixed per sample (\autoref{tab_parameters}). The variation is the interpolation scheme (TENO, WENO-Z). For the order independence, 5 configurations are also chosen by fixing the physical time ($t_0,t_1,t_2$), the resolution ($256\times 256$, $512\times 512$, $ 1024\times 1024$), the scheme (TENO or WENO-Z). The variation is done on the order (5,7). Besides different input variations,
this protocol is the same for testing H1 and H2.

The persistence curves are generated for all the samples. Then, we average the 5 persistence curves (one per solver) to obtain 2 average persistence curves with respect to the variable parameters (schemes or orders). Finally, we compute the difference of the integrals between the two averaged curves (grey area on \autoref{fig_gaussian_curve}). The small values on the curves under a persistence of $10^{-6}$ correspond to numerical noise coming from the different simulation steps. They are removed from the computation of the integral with a threshold at $10^{-6}$(\autoref{fig_gaussian_curve}.a). The integral curve difference corresponds to our metric allowing to precisely describe the similarity in the topology of the critical points. Bigger is the integral, the more different the topology of the flow is. Thus, to verify hypothesis H1 related to the scheme, we want the difference of the integrals to be high. To verify hypothesis H2 related to the orders, we want the difference of the integrals to be close to zero.

\begin{figure}
 \centering
 \vspace{-1ex}
 \includegraphics[width=\figureShrink\linewidth]{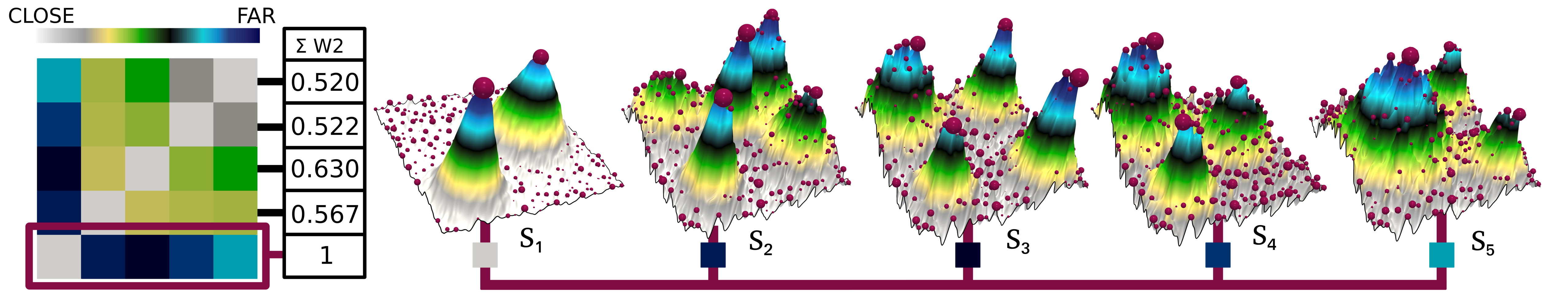}
 \mycaption{
 Wasserstein distance matrix for five inputs $S_1$, $S_2$, $S_3$,$S_4$, $S_5$
 generated respectively with two, five, four and three Gaussians
 with varying noise.
The sum of each
matrix line
is
normalized with respect to the scalar-field that maximizes the distances, here
$S_1$. We see that $S_1$ with only two Gaussians is very far from the other
datasets.}
 \label{gaussian_distance}
\end{figure}
\begin{figure}
\vspace{-1ex}
 \centering 
 \includegraphics[width=\figureShrink\linewidth]{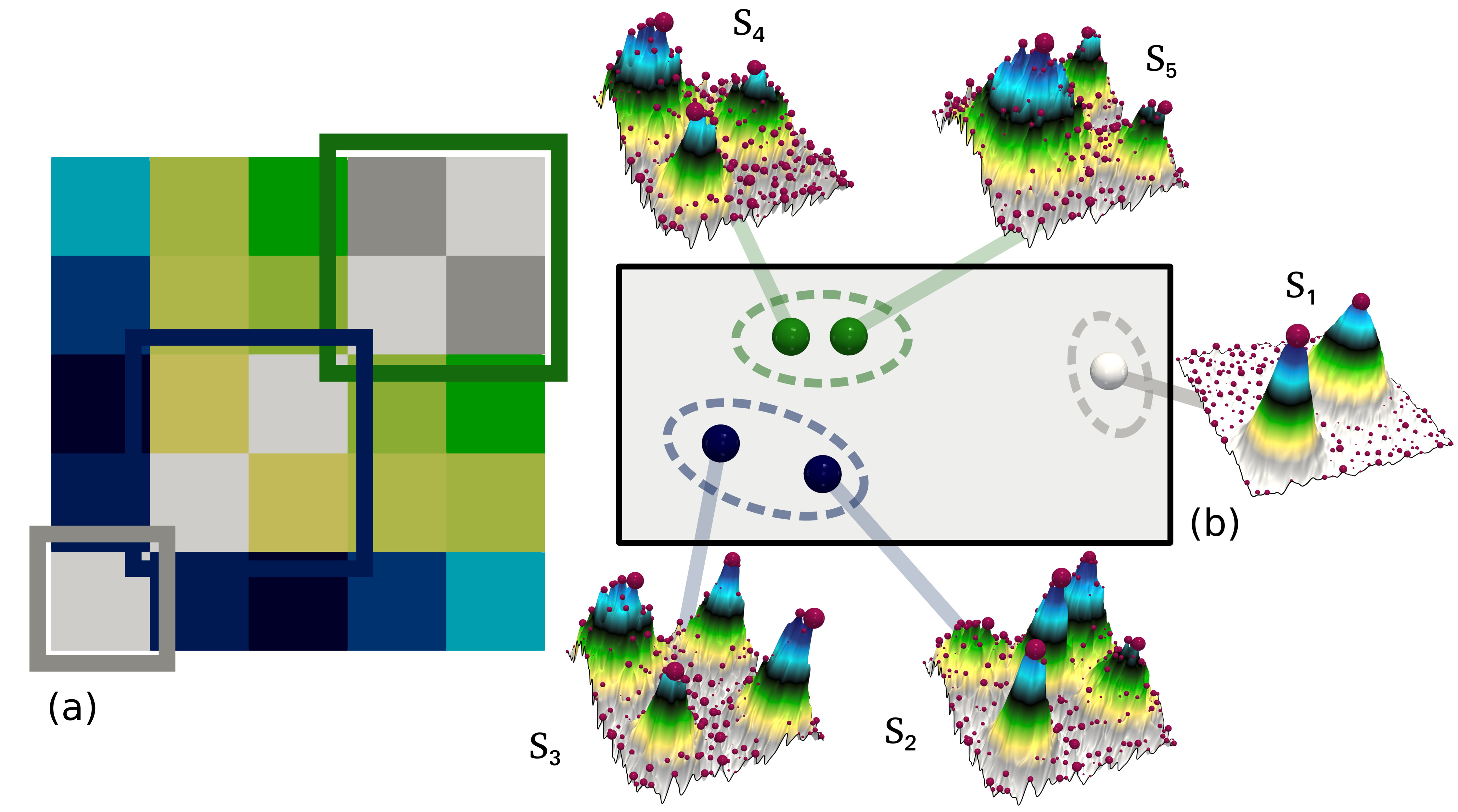}
 \mycaption{
 Wasserstein distance matrix for five inputs $S_1$, $S_2$, $S_3$, $S_4$, $S_5$ generated respectively with two, five, for and three Gaussian
functions with different noise levels. Point cloud of the inputs in the Wasserstein distance space colored according to the clusters obtained with the k-means clustering method. We can see that each terrain in a cluster has the same number of Gaussian and level of noise.}
 \label{gaussian_cluster}
\end{figure}
\begin{figure*}
 \centering 
 \includegraphics[width=\figureShrink\linewidth]{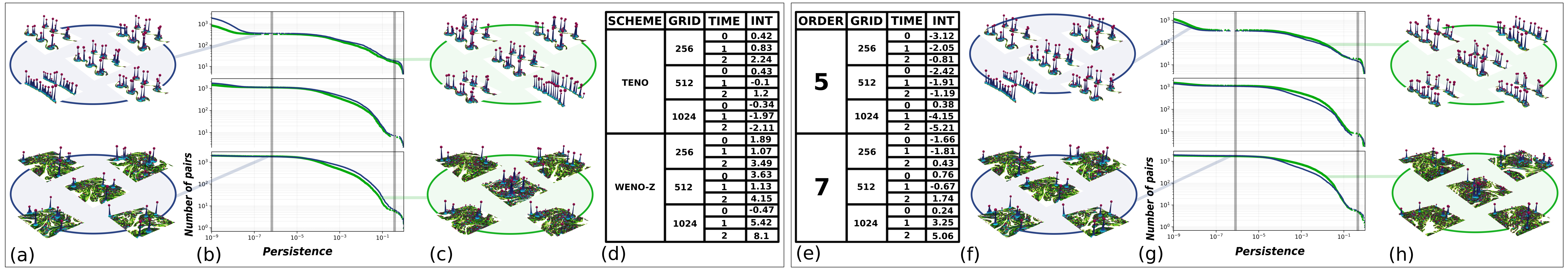}
 \mycaption{
Schemes (left) and order (right) studies.
Average
persistence curves for 5 configurations with variations of : a (WENO-Z,5), c
(WENO-Z, 7), f (WENO-Z, 5), h (TENO,5). (b,g) persistence curves at $t_0$ (top),
$t_1$ (middle), $t_2$ (bottom).
 Vertical lines on the curves correspond to critical points of small (left) and
high (right) persistence.
(d,e) Integral differences (grey area) between average persistence curves for
all variations.}
\vspace{2ex}
 \label{curv}
\end{figure*}

\begin{figure*}
 \centering
 \includegraphics[width=\figureShrink\linewidth]{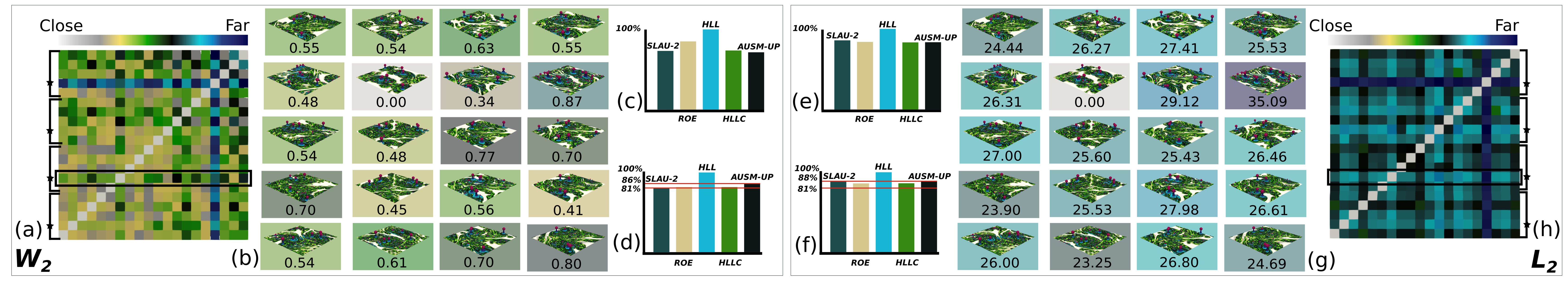}
 \mycaption{Comparison between the $\wasserstein{2}$ metric (left) and the standard $L_2$-metric (right) for isolating the HLL solver.
 (a,h) Distance matrix for 20 configurations at $t_2$ at $512 \times 512$. Black frames represent the distance between the TENO 7$_{th}$ order with the HLL (matrix lines marked by $\star$) and the other configurations (b,g). Histograms (c,e) are respectively the percentage average of the sum distance matrix of (a,h). Histograms (d,f) are respectively the percentage of the sum distance for all variations.}
 \label{fig_KHIdistance}
\end{figure*}

\subsection{Outlier distance profile}
\label{sec_outlier}
With this protocol (illustrated on toy examples,
\autoref{gaussian_distance}), we want to validate hypothesis H3 (\autoref{Hypotheses}), which means that for all the simulation configurations the HLL solver will be very different from other solvers to describe the Kelvin-Helmholtz instabilities. For this protocol, we take 5 simulation configurations where we fix the reconstruction (TENO or WENO-Z), the physical time ($t_0$, $t_1$, $t_2$), the mesh ($256\times 256, 512\times 512, 1024\times 1024$), the order (5 or 7) and we vary the solvers. The 5 different computations describing the same turbulent flow obtained with the solvers (HLL, SLAU2, AUSM$^+$-UP, HLLC and Roe) are analyzed regarding to the enstrophy.

A distance is used to compare the topology of the enstrophy. Many methods can be used to compute such a distance but in this protocol we focus on 2 metrics: the $L_2$-norm distance directly on the values of the enstrophy and the Wasserstein distance on the persistence diagrams. One can inject other distances if needed. For the Wasserstein, the saddle-maximum persistence diagram is computed on each result. Then, they are grouped in a unique dataset to compute a persistence diagram distance matrix (\autoref{gaussian_distance}). For the $L_2$-norm, a distance matrix  is also created where a line corresponds to the distance in the enstrophy field from one solver to the others.

Thus, the sum of the distances from one solver to the others is computed by
summing the distances on one line of the matrix. The total distance of one 
solver to the others, for all configurations, is simply the sum of all these sum 
distances for every line of the matrix which correspond to the same solver. We 
finally obtain one global distance per solver for all configurations. Finally 
the difference between the distance of the HLL and the distance of the maximizer 
(the second value if HLL is the maximum) gives a separation score. If the
difference is positive, then hypothesis H3 is verified whereas it is not if 
negative, because it means that another solver generates a flow topologically 
more different than the HLL. 
With this protocol, best separations are obtained for high absolute values.


\subsection{Unsupervised classification}
\label{sec_unsupervised}

With the last protocol (illustrated on toy examples in \autoref{gaussian_cluster}),
we want to validate hypotheses H4 and H5 (\autoref{Hypotheses}). We want to verify that the
simulations with the Roe and HLLC solvers are topologically close (hypothesis H4) and the
simulations with the AUSM$^+$-UP and SLAU2 solvers are topologically close (hypothesis H5). To do so, three clustering methods will be used based on Wasserstein distances and the L$_2$- norm(\autoref{sec_topology}).

For the first two clustering methods, we start by computing distance matrix with the protocol of the outlier distance profile \autoref{sec_outlier} using successively the Wasserstein distance and $L_2$-norm matrices (\autoref{gaussian_cluster}a). We apply a dimension reduction to project the distances of the matrix according to 2 components (\autoref{sec_topology}). This projection is used to generate clusters of the matrices with a k-means algorithm (\autoref{sec_topology}) as illustrated on \autoref{gaussian_cluster}b. The third clustering method uses directly the persistence diagrams(\autoref{sec_topology}) without using the distance matrix. All the persistence diagrams are merge into a single dataset to compute the Wasserstein distances between each diagram. The barycenter of persistence diagram is then used to directly compute a cluster, without dimension reduction, in the Wasserstein metric space \cite{vidal_vis19} with the ${W_2}$ distance. Then a k-means algorithm (\autoref{sec_topology}) is applied. 

With these three classification methods, we obtain different associations of our configurations. Each association is going to be scored with a measure of similarities between the clusters regarding to a reference cluster using the Rand Index \cite{rand1971objective}. This Rand index has a value between 0 and 1, with 0 indicating that two clusters do not agree on any pair of points and 1 indicating that the data clusters are exactly the same. Based on the properties of the solvers used in the simulation code HYPERION and detailed in \autoref{sec_solvers}, we define our reference cluster such that the first partition contains the AUSM$^+$-UP and SLAU2 solvers, the second partition the HLLC and Roe solvers and the third partition the HLL solver. The Rand Index is computed for each configuration and averaged per clustering method.
This
enables the
ranking of
the different solver
behaviors.
If the average Rand Index score is close to 1 then both hypotheses H4, showing similarity between the AUSM$^+$-UP and SLAU2 solvers and H5, showing the isolation of the HLL solver, are verified.

\section{Results}
\label{sec_experimentalResults}
This section presents our experimental results and their interpretations, for
the protocols presented in \autoref{sec_protocols}, applied on the ensemble data
described in \autoref{sec_caseStudy}
(publicly
available \cite{data}).

\subsection{Persistence curve study}

We applied protocol 1 using the persistence curves, on our ensemble dataset of Kelvin-Helmotlz instability (KHI) to verify the hypotheses of separation of the schemes (H1) and the independence of the orders (H2)(\autoref{Hypotheses}). 
The input parameters are setup as detailed in \autoref{sec_protocols}, generating 36 studies. The terrains and curves on
\autoref{curv} illustrate the result for one configuration with a 5th order WENO-Z (\autoref{curv}.a), a 7th order WENO-Z (\autoref{curv}.c), a 5th order WENO-Z (\autoref{curv}.f) and a 5th order TENO (\autoref{curv}.h). For the scheme comparison, most of the averaged persistence curves for the TENO schemes (blue curves on \autoref{curv}g) are above the WENO-Z curves (green curves on \autoref{curv}g).
The integral difference, between the average curves, obtain results between $[-5.6,5.2]$ (\autoref{curv}.e), which demonstrates differences on the topology of the enstrophy between the interpolation methods as expected.
Hypothesis H1 is verified on the KHI ensemble dataset. For the study on the 
independence of orders, we see that the averaged persistence curves are often 
close (\autoref{curv}b). However the integral differences obtained for this 
study show larger values for the WENO-Z, \emph{i.e} in between $[-0.5, 8.1]$ 
(\autoref{curv}.d). This analysis highlights that orders play a more important 
role, in terms of topology of the vortices, for WENO-Z than for TENO. Moreover, 
we observe that this difference tends to increase at $t_2$ for both studies 
confirming that the flow is composed of a larger number of vortex as the 
simulation evolves. Hypothesis H2 is verified for the TENO solvers but not for 
the WENO-Z
(\autoref{insights}).

\subsection{Outlier distance profile study}
\label{sec_KHIdistance}

To verify the HLL isolation states in hypothesis H3 (\autoref{Hypotheses}) on our ensemble dataset, we implemented our protocol 2 (\autoref{sec_protocols}) based on the Wasserstein distance and the $L_2$-norm (\autoref{sec_topology}). For this study we apply protocol 2 where the time and the resolution are fixed. The parameters that vary are the schemes ($\times 2$), the orders ($\times 2$) and the solvers ($\times 5$) (\autoref{tab_parameters}) thus generating 20 cases. All the distances have been computed according to the protocol of the outlier distance profile. These distances are represented by a global distance matrix where a line represents the 20 configurations (Wasserstein \autoref{fig_KHIdistance}.a and $L_2-$norm \autoref{fig_KHIdistance}.h) compared to a the HLL solver choosen as the reference. The matrix view of \autoref{fig_KHIdistance}b and \autoref{fig_KHIdistance}g show the KHI terrains and the distances of all configurations to the HLL solver.

The study has been done for all time steps and all resolutions generating nine
$20\times 20$ 
distance matrices, for each distance. The histograms (Figs. 
\ref{fig_KHIdistance}.c and \ref{fig_KHIdistance}.e) 
show the average 
of these nine distance matrices for the Wasserstein distance and the 
$L_2$-norm, expressed in terms of percentage according to the distance of HLL 
to the other solvers (HLL being the reference at 100\%). In this case the 
percentage difference in distances to HLL are about 18\% 
(\autoref{fig_KHIdistance}.d) for the Wasserstein and 13\% for the $L_2$ 
(\autoref{fig_KHIdistance}.f). These large percentages confirm that HLL is a 
solver that behaves differently from others.
As it does not take
into account contact discontinuities, the interfaces between the vortices
are much less defined than with the other solvers, resulting in a different
number of vortices. From a physical point of view, this result confirms the 
isolation of HLL in all cases. From a topological point of view, it shows that 
the Wasserstein distance is the best at differentiating the HLL solver from the 
others (the distance gap is always bigger than the $L_2$). For this large study 
of 18 distance matrices $20\times 20$, the hypothesis H3 is verified. 

\begin{figure*}
 \centering 
 \vspace{-1ex}
 \includegraphics[width=\figureShrink\linewidth]{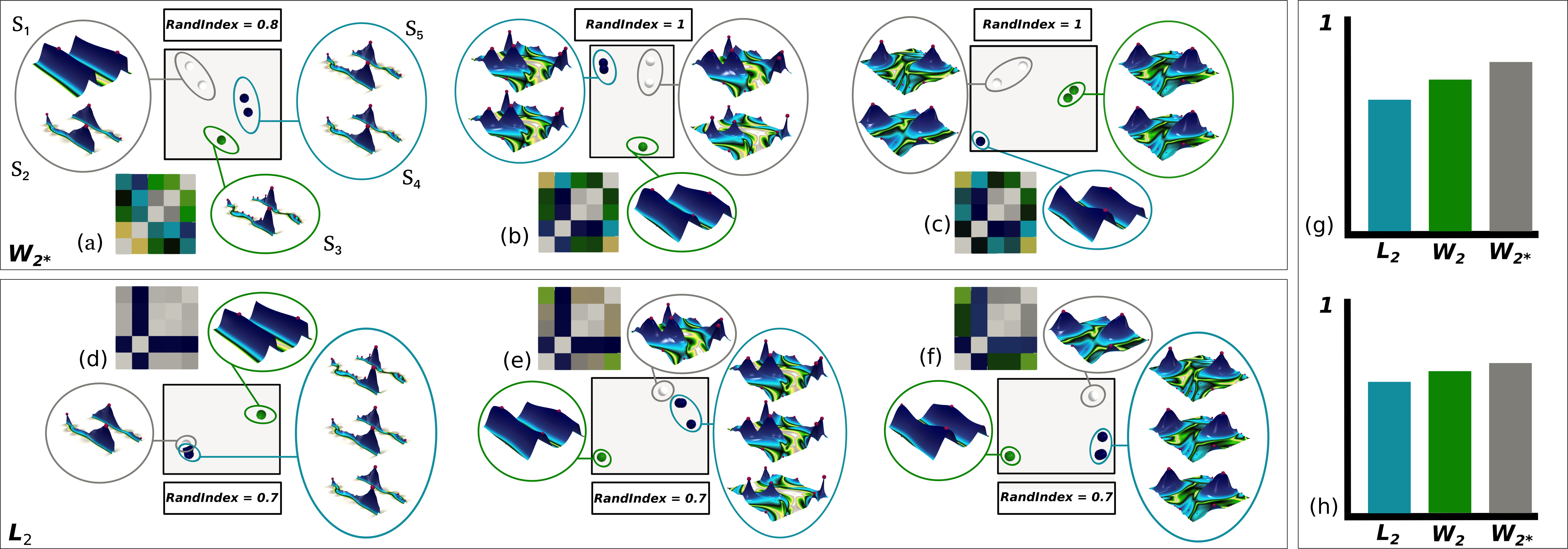}
 \mycaption{Comparison between the clustering on the Wasserstein metric space \cite{vidal_vis19} (top frame) and a clustering based on the traditional $L_2$ norm (bottom frame) for distinguishing FDS solvers from FTS solvers.
Point clouds at $t_0$, $t_1$,$t_2$ with a first order scheme at $256\times 256$.
The point cloud is a representation of the five scalar-fields in the distance
space colored according to the clusters obtained. The Rand Index are computed
with the five configurations $S_1$ (SLAU2), $S_2$ (HLL), $S_3$ (AUSM$^+$-UP),
$S_4$ (Roe), $S_5$ (HLLC). (g,h) Average Rand Index for all variations for the
high orders (bottom) and the first order (top).}
 \label{fig_khi_cluster}
\end{figure*}

\subsection{Unsupervised classification study}
 \label{sec_khi_cluster}
To improve our understanding on the behavior of the solvers into our simulation code, we implemented protocol 3 on the unsupervised classification (\autoref{sec_protocols}) to verified the hypotheses H4 and H5 (\autoref{Hypotheses}). The goal is to identify the separation of FDS type solvers from the FTS type solvers (\autoref{sec_solvers}). We are interested in the low Mach reconstructions (\autoref{sec_solvers}). The challenge comes from the fact that small vortices are reconstructed on only a few cells. So, we implemented protocol 3 with the distances and clustering method detailed in \autoref{sec_protocols} leading to 5 simulation configurations (\autoref{tab_parameters}) with variable solvers. To focus on the small vortices we used a threshold of 0.38 persistence for the topological methods. On the KHI ensemble, we generated 36 clusters from the threshold persistence diagrams and obtain the Rand Index for all of them. \autoref{fig_khi_cluster}.a, \autoref{fig_khi_cluster}.b, \autoref{fig_khi_cluster}.c show the $W_2*$ clustering for the three timesteps and \autoref{fig_khi_cluster}.d, \autoref{fig_khi_cluster}.e, \autoref{fig_khi_cluster}.f for the $L_2$.

Histogram \autoref{fig_khi_cluster}.h shows the average Rand Index for the 
three methods with a value of 0.63 for $L_2$, 0.66 for $W_2$ and 0.71 for $W_2*$.
There is very little difference between the topological and geometric results 
and each of the methods struggles to get the right cluster. Hypotheses H4 and H5 
are not verified for high orders. However, to highlight the differences between 
solvers, it is necessary to use a reference reconstruction that barely captures 
small scale turbulence due to order dissipation (\autoref{sec_solvers}).
Thus, we applied protocol 3 (\autoref{sec_protocols}) on a more restricted 
dataset at order 1. Histogram \autoref{fig_khi_cluster}.g shows the average 
Rand 
Index at order 1 with the three methods leading to 0.63 for $L_2$, 0.71 for $W_2$
and 0.78 for $W_2*$. In this case, we notice that for any reconstruction, the 
topological methods obtain better clustering. Moreover, the study with order 1 
shows that the $W_2*$ method enhances solver isolation.
With this
high score of the Rand Index hypotheses H4 and H5 are verified with the first 
order.


\subsection{Unanticipated insights}
\label{insights}

During the analysis of the persistence curves generated by our protocol 1, we found significant differences on the topology of the enstrophy between the orders for the WENO-Z. By increasing the order, we increase the accuracy of our calculation that generates more structures into the turbulent flow. On the other hand, there is no difference between the orders obtained with the TENO. This means that other ingredients in the TENO reconstruction play an important role in the computation of the turbulence such as the separation of the scales. In addition, the persistence curves also allowed us to observe that the WENO-Z schemes produce more numerical errors than the TENO. As presented in \autoref{sec_khi_cluster}, H4 and H5 hypotheses have not been verified for high orders. This means that the topological analysis does not capture the differences between the solvers. This may be due to the reconstructions which are accurate enough to calculate all velocities in the Kelvin-Helmholtz instability.


\subsection{Limitations}
As discussed in \autoref{sec_KHIdistance}, in comparison to the $L_2$ norm, the Wasserstein distance
improves
the separation of the HLL solver, but only by $5 \%$ (distance difference percentage). While this improvement may seem marginal, we would like to stress its significance given such challenging data,
in particular with regard to the traditional approach based on kinetic energy, shown \autoref{energie}, where the five solvers can hardly be distinguished from each other.
Similarly, we can see that the Rand Index score for the three clustering
methods detailed in \autoref{sec_khi_cluster} are quite close to each other as
illustrated on \autoref{fig_khi_cluster}.h. These close scores are due to the 
interpolations schemes (\autoref{sec_simulation}) which cover up the differences 
between the different solvers.
In other words, the variations in vortex distributions induced by the choice of solver are too subtle, given the importance of the interpolation order on the outcome. As shown in \autoref{fig_khi_cluster} (top), we were still able to overcome this limitation by considering
a
reconstruction that is not dedicated to turbulence, \emph{i.e.} an upwind scheme 
of order 1. This enabled us to exaggerate the impact of the solvers, thereby allowing us to
validate hypotheses H4 and H5 as reported in \autoref{sec_khi_cluster}.

\section{Conclusion}
In this paper, we have presented an experimental protocol for the comparison of
numerical methods on a Kelvin-Helmholtz instability using topological analysis.
An ensemble dataset of 180 members has been computed for this instability by a
simulation code developed in our institution and running on a supercomputer.
While traditional approaches based on the kinetic energy (\autoref{energie}) only enable
to validate the physical conformity of the generated flow,
our overall approach provides finer analyses. In particular,
the
protocol using the persistence curves (\autoref{sec_persistence}) allowed us to
observe differences between the TENO and WENO-Z reconstructions. It also
confirms an independence of the
reconstruction order (5 or 7) when
using the TENO scheme allowing
practical
computational speedup,
without loss of precision. The protocol
based on the Wasserstein distance (\autoref{sec_outlier}) succeeded in
discriminating the HLL solvers from other configurations, validating the use of
such a topological analysis to confirm domain field expectations. The last
protocol, based on recent clustering methods (\autoref{sec_khi_cluster})
successfully differentiates the topology of computations based on FDS (Flux
Difference Splitting) and FTS (Flux Type Splitting) solvers.
Overall, the validation of the hypotheses reported by CFD experts (\autoref{sec_hypotheses}) provides reliable indications for the tuning of a flow simulation, to help CFD users achieve the best balance between computation accuracy and speed.
%
%

The results obtained in this experimental study also show
the viability of topological methods for the representation and comparison of
Kelvin-Helmholtz
instabilities.
%
%
The interesting aspect of these topological protocols is that the
numerical method comparisons are based on physical differences rather than on
unreliable, low-level, pointwise measures.
The direction we wish to take now, for our future work, is
the extension of these protocols to 3D datasets of external hypersonics aerodynamics.
Another direction we
want to investigate is the evaluation of other tools used in the protocol such
as new topological distances \cite{pont_vis21} or clustering methods. Finally, this experimental
study allows us, with confidence, to consider applying these protocols to other
hydrodynamic turbulent flows studied in our institution in the domain of
hypersonic vehicle design.

\acknowledgments{
\vspace{-.5ex}
\small{
This work is partially supported by the European Commission  grant
ERC-2019-COG ``TORI'' (ref. 863464,
\href{https://erc-tori.github.io/}{https://erc-tori.github.io/}).
}
}



\clearpage
\bibliographystyle{abbrv-doi}
\bibliography{template}
\end{document}